# DEVELOPMENT OF A DTL QUADRUPOLE MAGNET WITH A NEW ELECTROFORMED HOLLOW COIL FOR THE JAERI/KEK JOINT PROJECT

K. Yoshino, E. Takasaki, F, Naito, T. Kato, Y. Yamazaki, KEK,
1-1 Oho, Tsukuba-shi, Ibaraki-ken, 305-0801, Japan
K. Tajiri, T. Kawasumi, Y. Imoto, Z. Kabeya, Mitsubishi Heavy Industries,
10 Oye-cho, Minato-ku, Nagoya, 455, Japan

*Abstract*

Quadrupole electromagnets have been developed with a hollow coil produced using an improved periodic reverse electroforming. These will be installed in each of the drift tubes of the DTL (324 MHz) as part of the JAERI/KEK Joint Project at the high-intensity proton accelerator facility. Measurements of the magnets' properties were found to be consistent with computer-calculated estimated. The details of the design, the fabrication process, and the measurement results for the quadrupole magnet are described.

## 1 INTRODUCTION

The research and development of focusing electromagnets for the 324-MHz DTL as part of the Japan Hadron Facility (JHF) started at KEK in 1996[1-3]. Since the operating frequency is much higher (324 MHz) than the conventional frequency (200 MHz), the size of the drift tube (DT) becomes smaller, resulting in many technical difficulties in fabricating a set of DT and quadrupole magnet (Q-magnet). Beam dynamics[2] require that the magnets and the DTs for the low-energy part of the DTL conform to the following specifications:

1. The magnetic field gradient must be variable.
2. The electromagnet must be installed within the compact DT (outer diameter within 140 mm, length about 52 mm).
3. The magnet must have a sufficiently large bore diameter (nearly 16 mm) and a high magnetic field (an integrated magnetic field gradient is 4.1 Tesla).
4. Expansion of the DT in the beam-axis direction should be less than 10 μm on one side during operation.
5. The deviation of a quadrupole field center from the mechanical center must be within 15 μm.

In order to satisfy the these requirements, the pulsed electromagnets have been selected, instead of permanent magnets. However, if we use the conventional hollow conductor type coil, it is very difficult to make an electromagnet which can be installed within a 324-MHz DT, since the rather large bending-radius is necessary for the hollow con-

Table 1: Design parameters of the Q-magnets and the DTs for the low-energy part of the DTL.

| | | |
|---|---|---|
| Magnet aperture diameter | (mm) | 15.6 |
| Core length : L | (mm) | 33.0 |
| Integrated field: GL(GLe) | (Tesla) | 4.1 |
| Effective length: Le | (mm) | 39.2 |
| Core material : | Silicon steel leaves 1) | |
| Main thickness of leaf | (mm) | 0.5 |
| Yoke outer diameter | (mm) | 115 |
| Nnumber of turns per pole | (turns/pole) | 3.5 |
| Maximum Ampere-Turns | (AT/pole) | 3500 |
| Excitation current ( Pulse ) | (A) | 780 |
| Pulse repetition rate | (Hz) | 50 |
| Pulse operation | rise time 5 ms, flat top duration 2 ms | |
| Minimum coil size | (mm) | h 5.5, w 5.3, t 1 |
| Voltage drop | (V) | 1.8 |
| Resistance | (mm Ω) | 2.3 |
| Inductance | (μH) | 18 |
| Water flow rate | (liter/min) | 1.0 |
| Water temperature increase | (°C) | 3.0 |
| Water pressure drop | (kg/cm$^2$) | 1.8 |
| DT outer diameter | (mm) | 140 |
| DT aperture diameter | (mm) | 13.0 |
| DT length | (mm) | 52.5 |

Note: 1) Nippon Steel Corporation., type 50H400[4]

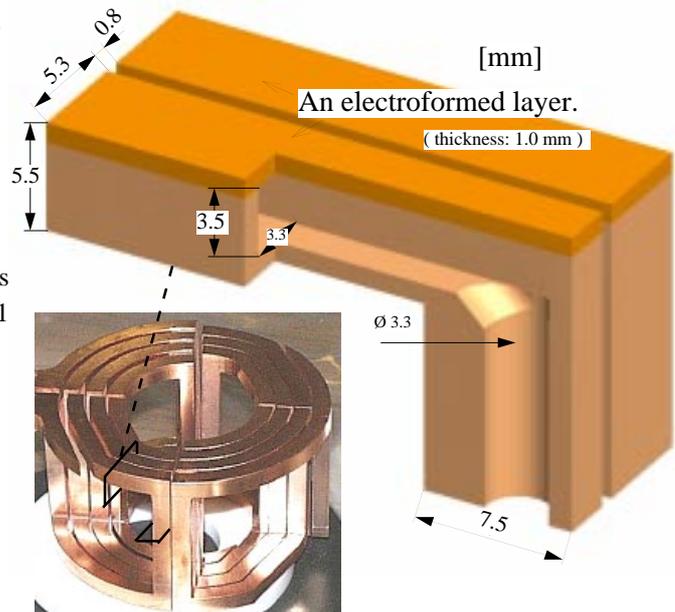

Figure 1: A detail of a corner part of the electroformed hollow coil.

ductors. Consequently, we newly developed an electroformed hollow coil. This new method makes use of an advanced Periodic Reverse (PR) copper electroforming[5,6] combined together with various kinds of machining processes without welding (except for the connection to the outside of DT), where the outer surfaces of the coil consists of an electroformed copper layer.

## 2 THE CHARACTERISTICS OF THE Q-MAGNETS AND THE DTS FOR THE LOW-ENERGY PART OF THE DTL

The design parameters of the Q-magnets and the DTs that we developed for the low-energy part of the DTL are listed in Table 1.

*2.1 Electroformed Hollow Coil*

Figure 1 shows a detail of a corner part of the electroformed hollow coil. The coil manufacturing process is outlined next. After cutting grooves for the water-cooling channel in a copper block (Fig. 2a), the grooves are filled with a wax, which is coated with silver powder to achieve electronic conductivity (Fig. 2b). A copper layer of 0.5 mm thickness is formed at each end face by PR copper electroforming (Fig. 2c). After machining the surfaces, additional copper deposits are formed by 0.5 mm thickness. After removing the wax and boring the pole-piece part (Fig. 2d), the coils of the end faces are separated using an end mill of 0.8 mm in diameter (Fig. 2e). The coils in the beam-axis direction are then separated to 1.0 mm by a wire-cutting machine (Fig. 1). In this way, the coil inside DT is molded without welding. Finally, the magnet leads are connected to the coils with silver brazing (Fig. 2f).

In order to reduce pressure drops and the effects of erosion, the water velocity in the coil is limited to under 2 m/s. For the same reason, a bending corner inside the water-cooling channel is partly cut. As a result, the measured pressure drops are 1.8 and 6.3 kg/cm$^2$ for the water flows of 1 and 2 Liter/minute, respectively.

*2.2 Q-Magnet and DT*

Some important properties of the magnets were measured before installing into the DT. Figure 3 shows the excitation-current dependences of the field gradients. The measured data are compared with those 3-D analyzed by MAFIA. Both are in agreement within approximately 2 %. Furthermore, the higher order multipole components in the magnetic field center measured by a rotating coil were sufficiently small, being less than 0.11% in comparison with the quadrupole component (Fig. 4). Also, the field center was deviated only by about 4 μm from the mechanical center.

After installing the magnet into the DT, some properties were also measured during field excitation. Figure 5 shows the dependence of the water-temperature increase upon the excitation-current. The temperature increase in the coil for water-flow rate of 1 liter/minute and the design excitation-current of 780 A was 3 ˚C, which is within the specification range. The drift tube was also water-cooled. However, variations in the flow rate of the DT has no measurable influence on the water-temperature of the coil. This is probably because the heat load on the coil is not so heavy.

The resonant frequency of the test tank[7] of 1.4 m was reduced by approximately 220 Hz, when the magnet was excited at the maximum current (for the design water-flow rate). This corresponds to an approximately 0.4 μm

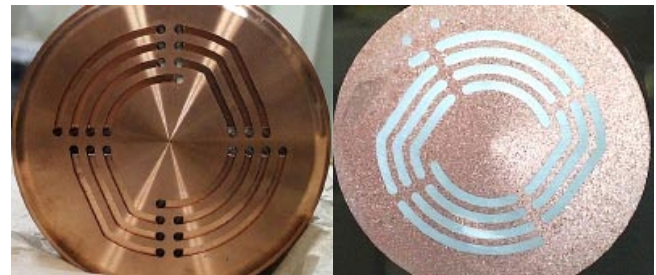

Fig. 2a: Groove processing for the water-cooling channel.

Fig. 2b: Filling with silver powder coated wax.

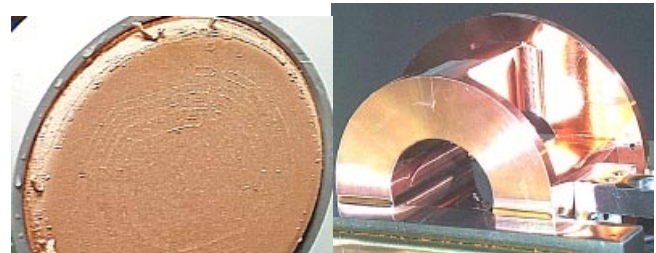

Fig. 2c: PR electroformed surface.

Fig. 2d: Cutting of the pole-piece part.

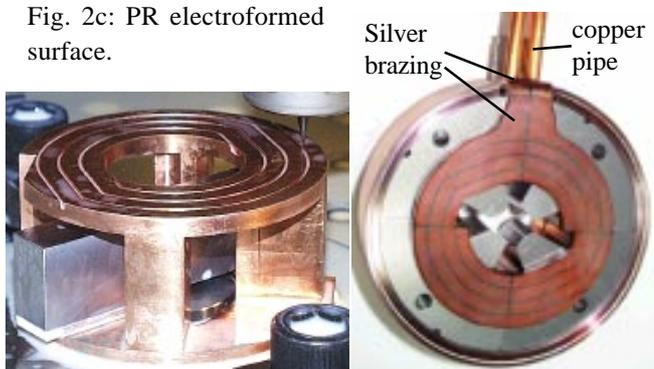

Fig. 2e: Separation of the coils by end mill.

Fig. 2f: Silver brazing of magnet leads.

Figure 2: Outline of the manufacturing process for the electroformed hollow coil.

expansion of the first drift-tube length.

## 3 THE SPECIFICATIONS OF ALL THE Q-MAGNETS

The specifications of all the Q-magnets in the DTL are shown in Table 2. Seven kinds of core lengths and five kinds of bore diameters are chosen in order to make trade-off between the requirements determined from the beam dynamics and the reduction in the fabrication cost. The cross sections of the coils for all the magnets are equal.

## 4 CONCLUSION

A prototype of the quadrupole electromagnet for 324-MHz DTL has been successfully made with the full specifications. The measured characteristics satisfied the requirements. In conclusion,
1. Quadrupole electromagnets have been developed with a hollow coil produced using an improved periodic reverse electroforming.
2. Measured field gradient agreed with the calculated one within approximately 2 %, and higher-order multipole components in the magnetic field center were sufficiently small, being less than 0.11% in comparison with the quadrupole component.
3. Since the pressure drop of the prototype coil is only 2 kg/cm$^2$ at the design water-flow rate, the electroformed coil can be adapted to those for the longer magnets.
4. The temperature increase in the coil at the design water-flow rate and excitation-current was 3 ˚C, which is within the specification.
5. The specifications for all the Q-magnets have been determined.

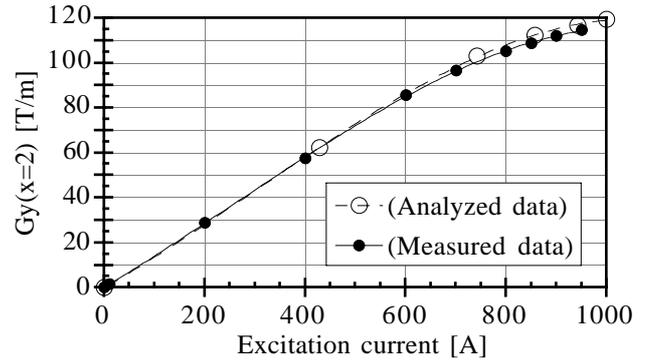

Figure 3: Comparison of excitation-current dependences on magnetic field gradient for measured data and 3-D analyzed data using MAFIA.

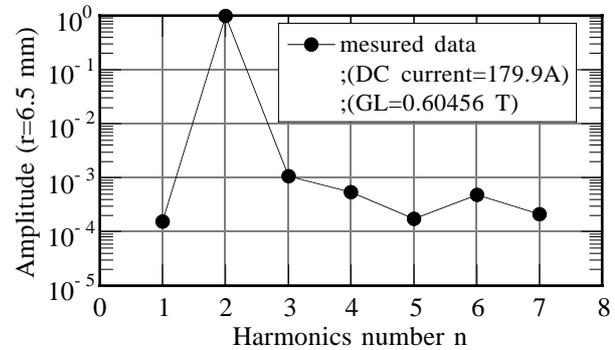

Figure 4: The higher-order multipole components in the center of the magnetic field.

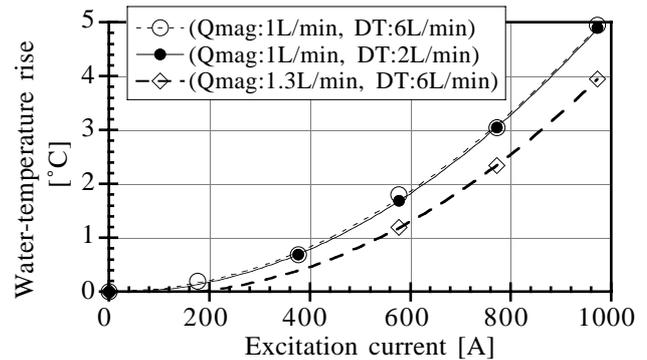

Figure 5: Excitation-current dependences on increase in the water-temperature of the electroformed coil.

Table 2: Specifications of all the Q-magnets in the DTL. (A unit is mm)

| DTL tank No. | | | | | | 1 | | 2 | 3 |
|---|---|---|---|---|---|---|---|---|---|
| Number of the DTs | | | | | | 77 | | 44 | 28 |
| Qmag outside diameter | | | | | | 115 | | | |
| Qmag core length | 33 | 35 | 50 | 76 | 80 | | 80 | 90 | 125 |
| Qmag bore diameter | 15.6 | 16 | 16 | 21 | 21 | | 25 | 25 | 29 |
| Number of the magnets | 6 | 17 | 33 | 1 | 20 | | 1 | 44 | 28 |